\pdfoutput=1 

\documentclass[preprint]{elsarticle}

\usepackage[T1]{fontenc} 
\usepackage{amsmath}    
\usepackage{graphicx}   
\usepackage{verbatim}   
\usepackage{color}      
\usepackage{hyperref}   
\usepackage{footnote}
\usepackage{slashed}
\usepackage{amsfonts}
\usepackage{amssymb}
\usepackage{adjustbox}  

\newcommand{\be}{\begin{equation}}
\newcommand{\ee}{\end{equation}}
\newcommand{\bea}{\begin{eqnarray}}
\newcommand{\eea}{\end{eqnarray}}
\newcommand{\nn}{\nonumber}

\journal{Nuclear Physics B}

\begin{document}

\title{\boldmath Testing quasi-Dirac leptogenesis through neutrino oscillations}

\newcommand{\affUFABC}{{\small \it Centro de Ci\^encias Naturais e Humanas\;\;\\
		Universidade Federal do ABC, 09.210-170,
		Santo Andr\'e, SP, Brazil}}
\newcommand{\affCU}{{\small \it Ottawa-Carleton, Institute for Physics, Carleton University,\\1125 Colonel By Drive, Ottawa, ON, K1S 5B6, Canada}}

\author[1]{C. S. Fong}
\ead{sheng.fong@ufabc.edu.br}
\author[2]{T. Gregoire}
\ead{gregoire@physics.carleton.ca}
\author[2]{A. Tonero}
\ead{alberto.tonero@gmail.com}

\address[1]{\affUFABC}
\address[2]{\affCU}

\begin{abstract}
	The lightness of the Standard Model (SM) neutrinos could be understood if their masses were to be generated by new physics at a high scale, through the so-called seesaw mechanism involving heavy fermion singlets. We consider a novel scenario where the new physics violates baryon minus lepton number by only a small amount resulting in both heavy fermion singlets and the SM neutrinos to split into pairs of quasi-Dirac states. Above the weak scale, the cosmic matter-antimatter asymmetry can be generated through resonant leptogenesis from decay of heavy singlets. Nontrivially, the CP violation for leptogenesis is bounded from above by the light neutrino mass splitting which can be probed in neutrino oscillation experiments.
\end{abstract}

\maketitle
\flushbottom

\section{Introduction}

The nature of the Standard Model (SM) neutrinos $\nu_L$, whether Dirac or Majorana is still an open question. In the former case, baryon number minus lepton number $B-L$ can remain an exact global symmetry while in the latter case, it has to be broken.

If neutrinos are Majorana particles with mass term $m_\nu \bar \nu_L \nu_L^c$, no new light degrees of freedom beyond the SM are required, and their lightness can be elegantly explained by the seesaw mechanism through the unique dimension-5 Weinberg operator \cite{Weinberg:1979sa}. Once the SM Higgs doublet acquires a vacuum expectation value (vev) $v = 174$ GeV, one obtains $m_\nu = c v^2/\Lambda$, where $c$ is some dimensionless coefficient and $\Lambda \gg v$ is the $B-L$-violating scale.

If neutrinos are Dirac particles, one will need to introduce new light degrees of freedom $\nu_R$'s (right-handed neutrinos) to couple to $\nu_L$ through $m_\nu \bar \nu_L \nu_R$. This Dirac mass term (protected by a $B-L$ symmetry) can arise at renormalizable level with $m_\nu = y_\nu v$ and the neutrinos' lightness is accommodated through a very tiny  
Yukawa coupling $y_\nu \sim 10^{-12}$. Another interesting possibility is to
have the neutrino Dirac mass suppressed by heavy $B-L$-conserving new physics scale $\Lambda$ through the Dirac seesaw mechanism. To realize this scenario  
some additional symmetry is needed to forbid the renormalizable mass term. 
For instance, in mirror world models \cite{Blinnikov:1982eh,Blinnikov:1983gh,Khlopov:1989fj} and Twin Higgs models \cite{Chacko:2005pe}, where the SM field content as well as gauge symmetry are duplicated, the new gauge symmetry  
forbids the renormalizable Dirac mass and the Dirac seesaw mechanism can be implemented. In this case, one has $m_\nu = c v f/\Lambda$ where $\nu_R$'s reside in the mirror lepton doublets and $f$ is the vev of the mirror scalar doublet.

The existence of new physics at a scale $\Lambda$ has important consequences for the generation of a baryon asymmetry through leptogenesis \cite{Fukugita:1986hr,Fong:2013wr}. In ref.~\cite{Earl:2019wjw}, it is shown that successful leptogenesis can be achieved in the mirror Dirac seesaw model \cite{Gu:2012fg,Earl:2019wjw} down to $10^7$ GeV, a scale still too high for experimental verification. In this work, we shall explore the \emph{quasi-Dirac} scenario \cite{Valle:1982yw} by introducing small $B-L$-violating terms to the model of \cite{Gu:2012fg,Earl:2019wjw}.  
As a consequence, light neutrinos split into quasi-Dirac active-sterile pairs\footnote{In \cite{Wolfenstein:1981kw,Petcov:1982ya} instead the authors consider active-active pairs of quasi-Dirac neutrinos, which have been ruled out by neutrino oscillation experiments~\cite{Frampton:2001eu,He:2003ih,Brahmachari:2001rn}.} where the mass squared splitting in the range $10^{-12} - 10^{-5} \,{\rm eV}^2$ can be constrained by neutrino oscillation experiments 
 which are sensitive to atmospheric and solar mass splitting ~\cite{Cirelli:2004cz,deGouvea:2009fp,Anamiati:2017rxw,Anamiati:2019maf}. At the same time, the heavy singlet fermions also split into quasi-Dirac pairs and CP violation in their decays can \emph{naturally} be enhanced to realize \emph{resonant} leptogenesis \cite{Pilaftsis:2003gt,Pilaftsis:2004xx,Pilaftsis:2005rv} around the weak scale as long as sufficient asymmetry is generated before electroweak (EW) sphaleron interactions become ineffective at $T \sim 132$ GeV as in the SM \cite{DOnofrio:2014rug}.\footnote{Leptogenesis where light neutrinos are also quasi-Dirac has been considered in ref.~\cite{Ahn:2016hhq}. However, in this work, $B-L$ is broken by a large Majorana mass term and the connection with low energy phenomena is lost.} Our main result is summarized by the following equation
\be
|\epsilon^{\rm max}| \simeq \frac{\delta m}{2m_\nu},
\ee
where $|\epsilon^{\rm max}|$ quantifies the maximal CP violation for leptogenesis while $\delta m$ is the small mass splitting of light neutrinos of mass scale $m_\nu$. Since successful leptogenesis put a lower bound on $|\epsilon^{\rm max}|$ while neutrino oscillation experiments can put an upper bound on $\delta m$,
this represents a rare testable leptogenesis model which is directly linked to low energy observable in neutrino oscillation phenomena. From minimality considerations, the observation of this small mass splitting would strongly suggest that neutrinos are indeed Majorana particles and allow to identify the parameter space which leads to viable leptogenesis.

\section{The model}
\label{sec:model}

In the Dirac mirror seesaw model, the SM and mirror sectors are connected through heavy Dirac singlet fermions $(N_{Ra}'^c,N_{Ra})$ \cite{Gu:2012fg,Earl:2019wjw}:
\bea\label{eq:lagr1}
{\cal L}&=&i\bar N_{Ra} \slashed{\partial} N_{Ra}+i\bar N'_{Ra} \slashed{\partial} N'_{Ra}
-\left(M_{ab} \bar N_{Ra}^c  N'_{Rb}+{\rm h.c.} \right)
\nn \\
&&
-\left(  y_{\alpha a}\bar{l}_{L\alpha}\tilde{\Phi}N_{Ra}
+ y'_{\alpha a}\bar{l'}_{L\alpha}\tilde{\Phi}'N'_{Ra}+{\rm h.c.} \right),
\eea
with $l_{L\alpha}$ and $\Phi$ the SM lepton and Higgs doublets charged under the SM EW $SU(2)_L \times U(1)_Y$ and $\tilde \Phi=i\sigma_2\Phi^*$ 
where $\sigma_2$ is the second Pauli matrix, 
while $l_{L\alpha}'$ and $\Phi'$ the mirror lepton and Higgs doublets that transform under the mirror EW group $SU(2)'_L \times U(1)'_Y$ and $\tilde \Phi'=i\sigma_2\Phi'^*$. Here we reserve $\alpha,\beta = e,\mu,\tau$ as the lepton and mirror lepton flavor indices and  $a,b=1,2,...$ the heavy fermion singlet family index. 

In this model, one can identify an anomaly-free global symmetry $U(1)_{\Delta_{\rm tot}}$ where~\cite{Earl:2019wjw}
\be 
\Delta_{\rm tot} \equiv (B-L)-(B'-L'),
\label{eq:Deltatot}
\ee
with $B$ ($L$) and $B'$ ($L'$) are respectively the baryon (lepton) number in the SM and the mirror sector with the following charge assignments:
\be 
\label{eq:L_charge}
\Delta_{\rm tot}(l_{L \alpha}) = \Delta_{\rm tot}(N_{Ra} ) = -\Delta_{\rm tot}(l'_{L\alpha})=-\Delta_{\rm tot}(N'_{Ra}). 
\ee
Though $\Delta_{\rm tot}$ is conserved, $B-L$ and $B'-L'$ are not separately conserved due to the mass term $M_a$. For this reason, it becomes possible to generate nonzero $B-L$ and $B'-L'$ asymmetries which remain equal in magnitude and \emph{sign} due to the conservation of $U(1)_{\Delta_{\rm tot}}$ \cite{Earl:2019wjw}. In this case, a nonzero CP violation requires at least two families of $N_R$ and $N_R'$. If this were not the case, one could make all the parameters real. 

The anomaly-free $U(1)_{\Delta_{\rm tot}}$ can be gauged and spontaneously broken in various ways. If the breaking happens at a scale $\mu \gtrsim M_i$, it is possible to generate small $\Delta_{\rm tot}$-violating terms by choosing the appropriate scalar field content under $U(1)_{\Delta_{\rm tot}}$~\cite{Earl:2019wjw}. In this work, we parametrize these small $\Delta_{\rm tot}$-violating terms as follows
\bea\label{eq:lagr2}
\slashed{\cal L}&=&
-\frac{1}{2}m_{ab} \bar N_{Ra}^c  N_{Rb}
-\frac{1}{2}m'_{ab} \bar N'^c_{Ra}  N'_{Rb}
\nn\\
&&
-\tilde y_{\alpha a}\bar{l}_{L\alpha}\tilde{\Phi}N'_{Rb}
-\tilde y'_{\alpha a}\bar{l'}_{L\alpha}\tilde{\Phi}'N_{Rb}+{\rm h.c.},
\eea
where $|m_{ab}|,|m'_{ab}| \ll M_a$, $|\tilde y_{\alpha a}| \ll |y_{\alpha a}|$ and $|\tilde y'_{\alpha a}| \ll |y'_{\alpha a}|$.
Now, not only are  $U(1)_{B-L}$ and $U(1)_{B'-L'}$  broken, $U(1)_{\Delta_{\rm tot}}$ is also broken. In the absence of $\Delta_{\rm tot}$ conservation, $B-L$ and $B'-L'$ asymmetries no longer have to be equal in magnitude and sign as we will explore next.

At this point it is useful to introduce two types of discrete transformations that interchange SM and mirror fields~\cite{Berezhiani:2003xm}. 
Let us first define the ``doubling'' transformation $Z_{2D}$ as the interchange of SM and mirror fields with same chirality $\psi_{L,R} \leftrightarrow \psi'_{L,R}$.   
The requirement of the lagrangian to be symmetric under $Z_{2D}$ will imply
\be 
M=M^T\qquad m = m' \qquad y = y' \qquad  \tilde y = \tilde y'\,.
\ee 
Let us also define a mirror parity, which we denote $Z_{2M}$, that interchanges SM fields and mirror fields with opposite chirality $\psi _{L,R} (t,\mathbf{x}) \to \psi'_{R,L} (t, -\mathbf{x})$. This can be considered as a generalization of parity and the requirement of the lagrangian to be symmetric under $Z_{2M}$ will imply
\be 
M=M^\dagger\qquad m^* = m' \qquad y^* = y' \qquad  \tilde y^*= \tilde y'\,.
\ee
These symmetries are motivated by generic mirror world models \cite{Blinnikov:1982eh,Blinnikov:1983gh,Khlopov:1989fj} and are used in explaining the coincidence in the amount of dark matter and baryon energy densities, and in addressing the little hierarchy problem in the context of Twin Higgs models \cite{Chacko:2005pe}. In the following we will consider the one family case and study in detail the connection between leptogenesis and neutrino masses, in these two symmetric limits.   

\section{Resonant leptogenesis and quasi-Dirac neutrinos}
\label{sec:resonant}
Due to new complex parameters in the $\Delta_{\rm tot}$-violating terms, even with one family of $N_{Ra}$ and $N_{Ra}'$, the CP violation for leptogenesis is nonzero.
In the following, we will consider the one family case ($a = 1$) and drop the index.\footnote{
	 From the leptogenesis point of view, the one-family analysis here can be seen as the limit of a multi-family model where the reheating temperature is of the order $T_{\rm RH} \sim M_1 \ll M_{a > 1}$ and resonant leptogenesis proceeds through the lightest ($N_{R1}, N'_{R1}$) where the contributions from heavier families will be subdominant. However, in order to explain the observed neutrino mass differences and mixing, at least one more family ($N_{R2}, N'_{R2}$) is needed.} 
	
Without loss of generality, in the basis where the charged SM and mirror lepton Yukawa matrices are real, positive and diagonal, we can choose $M$, $y_\alpha$ and $y_\alpha'$ to be real and positive. In principle, we can also make either $m$ or $m'$ real. Nevertheless, it is more convenient to work in the ``symmetric basis", where both $m$ and $m'$ are kept complex, because this will allow us to easily see that, if $Z_{2D}$ is exact, CP violation for leptogenesis vanishes identically.\footnote{
Starting from $Z_{2D}$ symmetric Lagrangian, if one makes either $m$ or $m'$ real, then the relation $\tilde y=\tilde y'$ will be spoiled as well, but in the end, the physics would remain the same.} 	
	
The mass matrix for the heavy fermion singlets in the basis $(N_R,N_R')$ reads
\bea
{\cal M} & = & \left(\begin{array}{cc}
m & M\\
M & m'
\end{array}\right).
\eea
We can diagonalize the symmetric mass term above by means of a unitary matrix $U$ such that $U^\dagger {\cal M} U^*  = {\rm diag}(M_1, M_2)\equiv \hat M $, with
\be
M_{1,2}^2 \simeq M^2\left(1\mp \frac{|m'+m^*|}{M} \right).
\ee
In term of the mass eigenstates $N_1$ and $N_2$ 
\be
\left(\begin{array}{cc}
	N_1 \\ N_2
\end{array}\right)
= U^T \left(\begin{array}{c}
	N_R \\ N'_R
\end{array}\right),
\ee
the relevant Lagrangian in the mass basis is rather compact
\be\label{eq:lagr_mass}
- ({\cal L} + \slashed{\cal L}) \supset
\frac{1}{2}M_i \bar N_i^c  N_i
+ Y_{\alpha i}\bar{l}_{L\alpha}\tilde{\Phi}N_i
+ Y'_{\alpha i}\bar{l'}_{L\alpha}\tilde{\Phi}'N_i
+{\rm h.c.},
\ee
where $Y_{\alpha i} =y_\alpha U_{1 i}^* + \tilde y_\alpha U_{2 i}^*$, $Y'_{\alpha i} = y'_\alpha U_{2 i}^* + \tilde y'_\alpha U_{1 i}^*$ with $i=1,2$.
The two pseudo-Dirac singlets can decay to the SM sector as well as the mirror sector $N_i \to l_\alpha \Phi, l'_\alpha\Phi'$ (and the CP conjugate states) with the tree-level total decay width given by
\be
\Gamma_i = \frac{M_i}{8\pi} \left[(Y^\dagger Y)_{ii} + (Y'^\dagger Y')_{ii}\right]. \label{eq:decay_width}
\ee
The CP violation in $N_i$ decays can be quantified by the following CP parameters
\bea
\epsilon_{i \alpha} &\equiv& \frac{\Gamma(N_i \to l_\alpha \Phi) - \Gamma(N_i \to \bar l_\alpha \bar \Phi)}{\Gamma_i}, \\
\epsilon'_{i\alpha} &\equiv& \frac{\Gamma(N_i \to l'_\alpha \Phi') - \Gamma(N_i \to \bar l'_\alpha \bar \Phi')}{\Gamma_i}.
\eea
Since we are considering resonant leptogenesis, the subdominant contributions from one-loop vertex corrections will be neglected and from the one-loop self-energy corrections, we have
\bea
\epsilon_{i\alpha} &=& \epsilon^c_{i\alpha} + \epsilon^v_{i\alpha}, \label{eq:ep_SM}  \\
\epsilon'_{i\alpha} &=& \epsilon'^c_{i\alpha} + \epsilon'^v_{i\alpha}, \label{eq:ep_SMp} 
\eea
where $\epsilon^c_{i\alpha}$ and $\epsilon'^c_{i\alpha}$ are the $\Delta_{\rm tot}$-conserving terms
\bea
\epsilon^c_{i\alpha} &\equiv& 
\frac{M_i}{(8\pi)^2\Gamma_i}
\sum_{j\neq i} \left\{
{\rm Im}[(Y'^\dagger Y')_{ij} Y^*_{\alpha i} Y_{\alpha j}] f_{ij}+ {\rm Im}[(Y^\dagger Y)_{ji} Y^*_{\alpha i} Y_{\alpha j}] g_{ij}
\right \}, \label{eq:CP_con}  \\
\epsilon'^c_{i\alpha} &\equiv& 
\frac{M_i}{(8\pi)^2\Gamma_i}
\sum_{j\neq i} \left\{
{\rm Im}[(Y^\dagger Y)_{ij} Y'^*_{\alpha i} Y'_{\alpha j}] f_{ij}+ {\rm Im}[(Y'^\dagger Y')_{ji} Y'^*_{\alpha i} Y'_{\alpha j}] g_{ij}
\right \}, \label{eq:CPp_con}  
\eea
and $\epsilon^v_{i\alpha}$ and $\epsilon'^v_{i\alpha}$ are the $\Delta_{\rm tot}$-violating terms
\bea
\epsilon^v_{i\alpha} &\equiv& 
\frac{M_i}{(8\pi)^2\Gamma_i}
\sum_{j\neq i} \left\{
{\rm Im}[(Y^\dagger Y)_{ij} Y^*_{\alpha i} Y_{\alpha j}] f_{ij}+ {\rm Im}[(Y'^\dagger Y')_{ji} Y^*_{\alpha i} Y_{\alpha j}] g_{ij}
\right \} \label{eq:CP_vio},  \\
\epsilon'^v_{i\alpha} &\equiv& 
\frac{M_i}{(8\pi)^2\Gamma_i}
\sum_{j\neq i} \left\{
{\rm Im}[(Y'^\dagger Y')_{ij} Y'^*_{\alpha i} Y'_{\alpha j}] f_{ij}+ {\rm Im}[(Y^\dagger Y)_{ji} Y'^*_{\alpha i} Y'_{\alpha j}] g_{ij}
\right \} \label{eq:CPp_vio} .
\eea
The regulated one-loop functions~\cite{Pilaftsis:2003gt} are given by $f_{ij} \equiv \frac{\sqrt{x_{ji}}(1 - x_{ji})}{(1-x_{ji})^2 + a_{ji}} $ and $g_{ij} \equiv \frac{1 - x_{ji}}{(1-x_{ji})^2 + a_{ji}} $, with $x_{ji} \equiv M_j^2/M_i^2$ and $a_{ji} \equiv \Gamma_j^2/M_i^2$. 
Resonant enhancement occurs when the mass splitting is of the order of the decay width and the maximum value is achieved for $(1 - x_{ji})^2 = a_{ji}$, 
or in terms of the Lagrangian parameters, the resonant condition reads
\be
|m' + m^*| \simeq \frac{\Gamma}{2} = \frac{M}{32\pi} \left(y^2 + y'^2\right), \label{eq:res_con}
\ee
where $y^2 \equiv \sum_\alpha y_\alpha^2$ and $y'^2 \equiv   \sum_\alpha y_\alpha'^2$ and $\Gamma \equiv (y^2+y'^2)M/(16\pi)$ is obtained from eq.~\eqref{eq:decay_width} keeping only the leading term.\footnote{One-loop radiative corrections to the Majorana mass parameters are
	\be
	\delta m\sim \frac{2M}{(4\pi)^2}\left( y_\alpha \tilde y_\alpha^* + y'_\alpha \tilde y'_\alpha \right),
	\quad \delta m'\sim\frac{2M}{(4\pi)^2}\left( y'_\alpha \tilde y'^*_\alpha + y_\alpha \tilde y_\alpha \right),
	\ee 
    and therefore for $|\tilde y_\alpha| \ll y_{\alpha}$ and $|\tilde y'_{\alpha}| \ll y'_{\alpha}$ the resonant condition is not spoiled by those corrections.
}
Notice that the $\Delta_{\rm tot}$-conserving terms in eq. \eqref{eq:CP_con} and \eqref{eq:CPp_con} vanish upon summing over all the final states, i.e. $\sum_\alpha (-\epsilon^c_{i\alpha}+\epsilon'^c_{i\alpha})=0$, whereas for the $\Delta_{\rm tot}$-violating terms  in eq. \eqref{eq:CP_vio} and \eqref{eq:CPp_vio}, we have that $\sum_\alpha(-\epsilon^v_{i\alpha} +\epsilon'^v_{i\alpha}) \neq 0$.
When taking $i \leftrightarrow j$ in eqs.~\eqref{eq:CP_con}--\eqref{eq:CPp_vio}, the imaginary part of the couplings change signs while $f_{ij} = -f_{ji}$ and $g_{ij} \approx - g_{ji}$ (the difference is linear in $\Delta_{\rm tot}$-violating parameters). Hence at the leading order in $\Delta_{\rm tot}$-violating parameters, we have $\epsilon_{1\alpha} = \epsilon_{2\alpha}$ and  $\epsilon'_{1\alpha} = \epsilon'_{2\alpha}$.

In the $Z_{2D}$ symmetric case we have $y_\alpha = y_\alpha'$ and $\tilde y_\alpha = \tilde y'_\alpha$ and the CP parameters in eqs.~\eqref{eq:ep_SM} and \eqref{eq:ep_SMp} are given by
\be
\epsilon_{1\alpha}  = -\epsilon'_{1\alpha} = \frac{M_1}{(8\pi)^2\Gamma_1}
y_\alpha^2{\rm Re}(w)  ( f_{12}+g_{12} )
 \sin\phi, \label{eq:ep_Z2} 
\ee
and
\be 
\epsilon_{1}  =-\epsilon'_{1}  
= \frac{M_1}{(8\pi)^2\Gamma_1}
  y^2{\rm Re}(w)  \left( f_{12}
 +  g_{12}
 \right)\sin\phi\,. \label{eq:eptot1_Z2} 
\ee
where $\phi \equiv \arg(m' + m^*)$, $w_\alpha \equiv y_\alpha \tilde y_\alpha$ and $w \equiv \sum_{\alpha} w_\alpha$.
Since in the $Z_{2D}$ limit we have that $m=m'$, this implies $\phi=0$ and the CP parameters in the equations above vanish identically. This can also be seen by the fact that in the $Z_{2D}$ limit $Y'_{\alpha 1}=-Y_{\alpha 1}$ and $Y'_{\alpha 2}=Y_{\alpha 2}$ at leading order in $\Delta_{\rm tot}$  breaking and substituting these relations into eq~\eqref{eq:CP_con} and eq.~\eqref{eq:CP_vio}. In the following we consider some $Z_{2D}$ breaking by allowing $\phi\neq 0$.
The sum of CP parameters are vanishing for each flavor $\epsilon_{1\alpha} + \epsilon'_{1\alpha} =0$ while $-\epsilon_{1\alpha} + \epsilon'_{1\alpha} = -2 \epsilon_{1\alpha} \neq 0$, as expected.
This implies that the SM asymmetry in the flavor charge $\Delta_\alpha \equiv \frac{B}{3}-L_\alpha$ will have the same magnitude and opposite in sign to the mirror $\Delta'_\alpha \equiv \frac{B'}{3}-L'_\alpha$ flavor charge asymmetry such that the total asymmetry in $\Delta_{\rm tot} = \sum_\alpha (\Delta_\alpha - \Delta'_\alpha)$ is nonvanishing.
At resonance fulfilling $(1 - x_{ji})^2 = a_{ji}$, the maximal CP parameters summing over flavor $\alpha$ are
\be \label{eq:CPmax_ZD}
|\epsilon_{1\alpha}^{\rm max}| = |\epsilon_{1\alpha}^{'\rm max}| 
\simeq  
\frac{y_\alpha^2\left| {\rm Re}(w) \right|}{ (y^2)^2}
\qquad
{\rm and}
\qquad 
|\epsilon_{1}^{\rm max}| = |\epsilon_{1}^{'\rm max}| 
\simeq  
\frac{\left| {\rm Re}\,w \right|}{y^2} \,.
\ee

In the $Z_{2M}$ symmetric case, we have  $y_\alpha^* = y_\alpha'$ and $\tilde y_\alpha^* = \tilde y'_\alpha$ and the CP parameters in eqs.~\eqref{eq:ep_SM} and \eqref{eq:ep_SMp} are given by
\be
\epsilon_{1\alpha}  = -\epsilon'_{1\alpha} = -\frac{M_1}{(8\pi)^2\Gamma_1}
y_\alpha^2 {\rm Im}(\,w\, e^{-i\phi})  ( f_{12}+g_{12} )
\ee
and
\be 
\epsilon_{1}  = -\epsilon'_{1} = -\frac{M_1}{(8\pi)^2\Gamma_1}
y^2 {\rm Im}(\,w\, e^{-i\phi})  ( f_{12}+g_{12} ) \,.
\ee
In this case $\phi \equiv \arg(2 m^*)$, the CP violation is nonzero in general and its maximal value is given by
\be\label{eq:CPmax_ZM}
\left|\epsilon_{1}^{\rm max}\right| = |{\epsilon'_1}^{\rm max}|
\simeq \frac{|w|}{y^2} .
\ee

Next, after the EW and mirror EW symmetries are broken, from eq.~\eqref{eq:lagr_mass}, we can write down the mass term for the light neutrinos using the seesaw formula in the basis $(\nu_L, \nu'_L)$
as a $6 \times 6$ matrix of rank-2 
\be
M_\nu = - m_D \hat M^{-1} m_D^T \,,\label{eq:seesaw}
\ee
where $m_D$ is given by the following $6 \times 2$ matrix 
\be
m_D \equiv
\left(
\begin{array}{c}
\upsilon Y  \\
f Y' 
\end{array}
\right)\, .
\ee

The mass matrix in eq.~\eqref{eq:seesaw} can be diagonalized by means of a unitary transformation $U_\nu$ such that  $U_\nu^\dagger M_\nu U_\nu^*  = {\rm diag}(m_-, m_+,0,0,0,0)$, where the positive mass eigenvalues are 
\be
m_\mp = m_\nu \mp\delta m,
\ee
with
\bea
m_\nu & \equiv & \frac{y y'vf}{M}, \quad
\delta m \simeq \Bigg| \frac{w^* v^2 + w' f^2 }{M} \Bigg|\,. \label{eq:light_nu_splitting}
\eea
In the equation above, we denote $y=\sqrt{y^2}$, $y'=\sqrt{y'^2}$ and in $\delta m$, we have neglected $m/M$ and $m'/M$ terms that, in the parameter space we are considering, are very small due to the resonant condition of eq.~\eqref{eq:res_con}.

In the $Z_{2D}$ and $Z_{2M}$ symmetric limit, we have $\delta m=2|{\rm Re}\, w|\upsilon^2/M$ and $\delta m=2|w|\upsilon^2/M$, respectively. Therefore from eqs.~\eqref{eq:CPmax_ZD} and \eqref{eq:CPmax_ZM} we obtain the following intriguing relation which connects high scale parameters responsible for leptogenesis with low energy observable
\be
\left|\epsilon^{\rm max}_1\right| \simeq \frac{\delta m}{2m_\nu}.
\label{eq:epmax_Z2}
\ee
To have successful leptogenesis with a certain $\epsilon^{\rm max}_1$, one can derive the allowed quasi-Dirac mass splitting in the light neutrino as 
\be
\delta m \gtrsim 2\times 10^{-8}\,{\rm eV} \frac{m_\nu}{0.1\,{\rm eV}} \frac{\left|\epsilon^{\rm max}_1\right|}{10^{-7}}.
\ee
The mass splitting above falls in the range which can be tested in solar and atmospheric neutrino oscillation experiments. In general, these experiments can put an upper bound on $\delta m$, which allows to completely cover the regime of viable leptogenesis. As a comparison, while the Davidson-Ibarra bound \cite{Davidson:2002qv} for type-I seesaw provides a minimum value for the heavy singlet mass scale, usually far beyond the scale reachable in experiments, the quasi-Dirac model provides a lower bound for $\delta m$ directly testable in experiments.

\section{Results and discussions}
\label{sec:discussion}
 
To complete the discussion of leptogenesis, we have to consider the $\Delta_{\rm tot}$-violating washout terms from the following scattering processes $l_\alpha \Phi \leftrightarrow l'_\beta \Phi'$, $l_\alpha \Phi \leftrightarrow \bar l_\beta \bar \Phi$, $l'_\alpha \Phi' \leftrightarrow \bar l'_\beta \bar \Phi'$, and the $\Delta_{\rm tot}$-conserving ones from $l_\alpha \Phi \leftrightarrow \bar l'_\beta \bar \Phi'$, $l_\alpha \Phi \leftrightarrow l_\beta \Phi$, $l'_\alpha \Phi'\leftrightarrow l'_\beta \Phi'$.
For $\Delta_{\rm tot}$-violating processes, at leading order, the amplitude from the exchange of $N_1$ cancels the amplitude from the exchange of $N_2$ up to a term linear in $\Delta_{\rm tot}$-violating parameters.\footnote{This cancellation was first pointed in ref.~\cite{Blanchet:2009kk} by arguing that the interference term in cross section is crucial for cancellation (see also refs.~\cite{Blanchet:2010kw,Agashe:2018cuf}). In fact, it is more direct to see this cancellation at the level of amplitude.} We consider the scatterings when the intermediate particles $N_1,N_2$ are \emph{on-shell} and neglect off-shell contributions which are of higher order in the Yukawa couplings. 
From explicit calculations, we found that all the $\Delta_{\rm tot}$-violating scattering processes come with an additional factor $|m'+m^*|^2/(2\Gamma^2)$ with respect to the $\Delta_{\rm tot}$-conserving processes, which at resonance amounts to a suppression by $|m'+m^*|^2/(2\Gamma^2)\sim 1/8$ (see  eq.~\eqref{eq:res_con}).\footnote{The factor is not valid when $|m'+m^*| \gtrsim \Gamma$ i.e. when the quasi-Dirac $N_i$ no longer overlap to lead to the cancellation discussed in ref.~\cite{Blanchet:2009kk}.} Hence, the flavor-violating but $\Delta_{\rm tot}$-conserving processes are generically as important as the $\Delta_{\rm tot}$-violating ones. In the regime when the $N_i$ decay rate is much faster than the Hubble rate, as long as the $y$ and $y'$ are not extremely hierarchical among different flavors, flavor equilibration in which asymmetries are equally distributed among all the flavors, will be achieved dynamically, \emph{independently} of the flavor structure of the CP parameters. Flavor equilibration can also be enforced by hand in all regime if the $N_i$ decay branching ratios as well as the CP parameters are equal in all flavors. 

\begin{figure}[t]
	\centering
    \includegraphics[width=0.9\columnwidth]{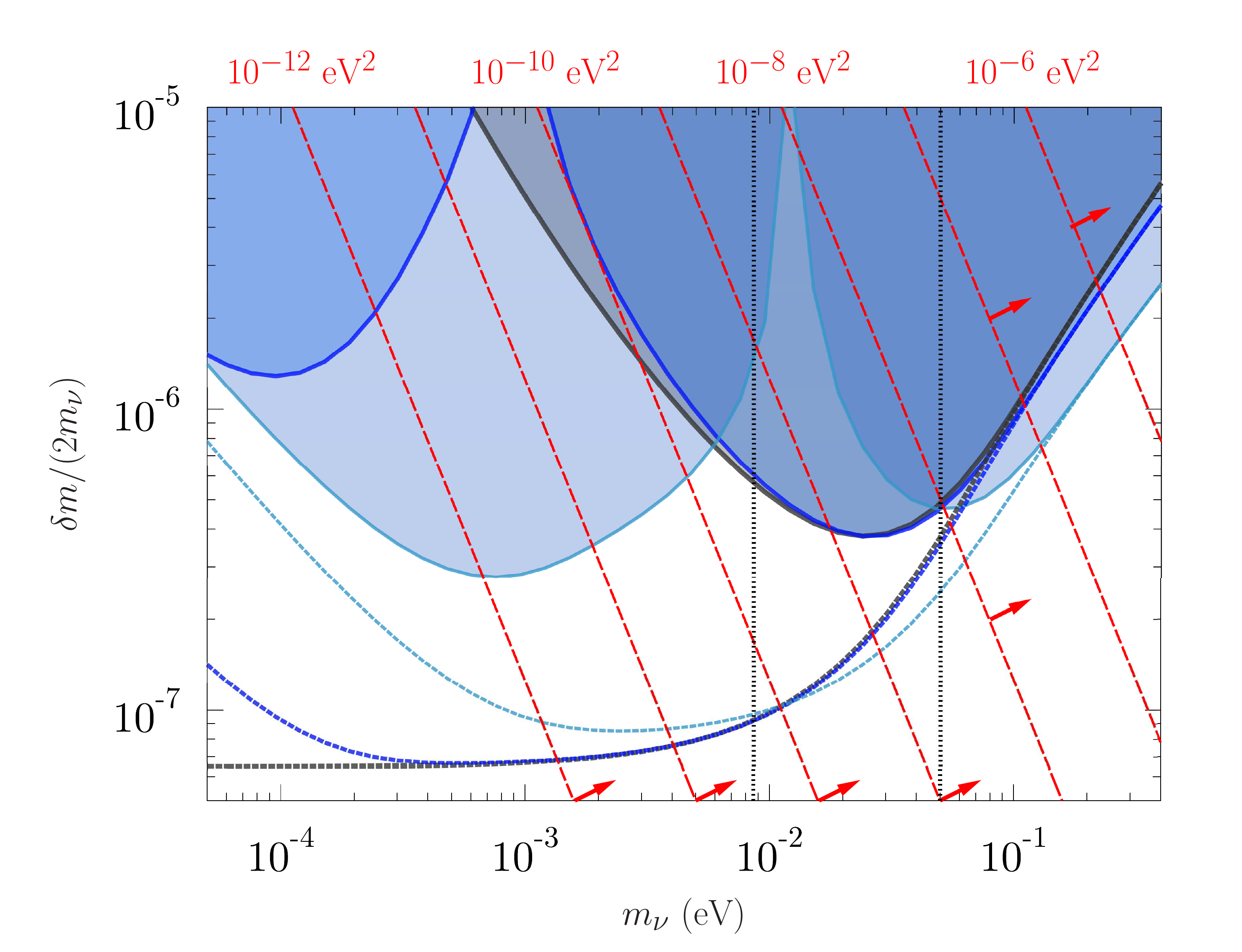}
    \caption{Regions in the $m_\nu$ vs $\delta m/(2 m_\nu)$ plane where sufficient baryon asymmetry can be generated for 
   	$M \gg 1\,{\rm TeV}$ (gray), $M=1\,{\rm TeV}$ (blue) and $M=500\,{\rm GeV}$ (light blue) for zero (solid) and thermal (short dashed) initial $N_i$ abundance. Long dashed red lines indicate parameter space which can be constrained in neutrino oscillation experiments (see text for details). 
   	The two vertical black dotted lines indicate respectively the scale of the solar $m_{\rm sol} = 8.6$ meV and the atmospheric mass splitting $m_{\rm atm} = 50$ meV.
   		\label{fig:plot}}
\end{figure}

We consider $Z_{2D}$ and $Z_{2M}$ symmetric scenarios as our benchmark scenarios and assume flavor equilibration. 
These two benchmarks lead to similar results as in both scenarios the CP violation parameter is bounded by eq. \eqref{eq:epmax_Z2}.
Under the above assumptions and setting the CP violation to the maximum value as in eq. \eqref{eq:epmax_Z2}, the final baryon asymmetry (including EW sphalerons processes) can be determined in terms of $M$, $m_\nu$ and $\delta m$ from solving the Boltzmann equations taking into account the scattering processes discussed above.
In Figure \ref{fig:plot}, we plot the regime where sufficient baryon asymmetry is generated in the plane of $m_\nu$ and $\delta m/(2 m_\nu)$. The result in Figure \ref{fig:plot} holds for both $Z_2$ scenarios.
As references, the two dotted vertical black lines indicate the solar $m_{\rm sol}=8.6$ meV and atmospheric $m_{\rm atm}=50$ meV mass scales. The gray, blue and light blue solid lines represent the parameter space where the observed baryon asymmetry is obtained for $M \gg 1$ TeV, $M = 1$ TeV and $M = 500$ GeV respectively, with zero initial $N_i$ abundance. Within the shaded areas, the baryon asymmetry is above the observed value. For the case of $M = 1$ TeV and $M = 500$ GeV, the parameter space is separated into two islands due to sign change in the final baryon asymmetry towards small $m_\nu$ as not all $N_i$ can decay before the EW sphaleron processes freeze out.
The short dashed lines with the same color coding are for thermal initial $N_i$ abundance where above the lines, the baryon asymmetry is above the observed value.
The red dashed lines indicate
the mass squared difference of quasi-Dirac light neutrinos\footnote{A nice feature of this model is that since leptogenesis proceeds within one family, this will connect leptogenesis to a specific generation of light neutrino eigenstates and its mass splitting.}
\be
\varepsilon^2 \equiv 4m_\nu \delta m,
\ee
ranging from $10^{-12}\,{\rm eV}^2$ to $10^{-6}\,{\rm eV}^2$. The arrows represent the parameter space which can potentially be excluded in neutrino oscillation experiments.
Solar neutrino experiments are not sensitive to values of $\varepsilon^2 \lesssim 10^{-12}\,{\rm eV}^2$,  but this could by probed by measuring the flavor content of high-energy astrophysical neutrinos \cite{Beacom:2003eu,Esmaili:2009fk,Esmaili:2012ac,Joshipura:2013yba}. The neutrino oscillation constraints on $\varepsilon$ depend on which light neutrino mass eigenstate $m_k$ ($k=1,2,3$) is split (denoting the splitting by $\varepsilon_k^2$). In ref.~\cite{Anamiati:2017rxw}, a two-parameter fit was performed (turning on one $\varepsilon_k^2$ and another new mixing angle at a time), leading to constraints in the range $\varepsilon_k^2 \lesssim  10^{-12} - 10^{-5}\,{\rm eV}^2$ for $k=1,2$, 
 where the strongest constraints come from solar neutrino data in Super-K and Borexino. Larger values of $\varepsilon_k^2$ are also allowed for fine-tuned values of the mixing angle. For $k=3$, the bound is in general much weaker~\cite{Anamiati:2017rxw}: $\varepsilon_3^2 \lesssim 10^{-5}\,{\rm eV}^2$ from Super-K, DayaBay, MINOS and T2K.
As we can see in Figure \ref{fig:plot}, the parameter space that is being probed by solar and atmospheric neutrino oscillation overlaps with the one where leptogenesis is viable

In our model, lepton flavor violation involving charged leptons can be induced at one-loop due to the heavy quasi-Dirac fermions $N_i$. Let us focus on $\mu \to e \gamma$ with the current experimental bound of ${\rm Br}(\mu \to e\gamma) < 4.2 \times 10^{-13}$ \cite{TheMEG:2016wtm}. In both $Z_{2D}$ and $Z_{2M}$ symmetric cases with flavor equilibration, we obtain
\bea
{\rm Br}(\mu \to e\gamma) & \approx & 6 \times 10^{-27}
\left(\frac{G_\gamma^2(M^2/M_W^2)}{0.7}\right) \left(\frac{m_\nu}{0.1\,{\rm eV}}\right)^2
\left(\frac{0.5\,{\rm TeV}}{M}\right)^2, 
\eea
where $M_W$ is the $W$ boson mass and $G_\gamma(x)$ is the loop function given by \cite{Ilakovac:1994kj}, 
with $G_\gamma^2(M^2/M_W^2) \approx 0.7$ for $M = 0.5$ TeV. This is far below the current experimental bound. 

Next, at leading order in $\Delta_{\rm tot}$-breaking parameters, the neutrinoless beta decay rate in our model is proportional to 
\be
(M_\nu)_{ee} \simeq -\frac{2 y_e \tilde y_e v^2}{M}\,.
\ee
Therefore the rate is suppressed by $\tilde y_e \ll y_e$. Even with $Z_2$ breaking, $y_e \gg y'_e$ and assuming that $y_e$ is the dominant one, we have from eq.~\eqref{eq:light_nu_splitting}, $|(M_\nu)_{ee}| \approx 2\delta m$ \cite{Valle:1982yw}. Hence this observable is not likely to be measured even in the next generation experiments with sensitivity reaching $(M_\nu)_{ee} \sim 10\,{\rm meV}$ \cite{Agostini:2017jim}. 

A direct verification of quasi-Dirac leptogenesis would be through the production of the heavy singlet fermions at particle colliders with the measurements of the ratio of same-sign to opposite sign dileptons in their decays $W \to \ell N \to \ell \ell j j$ \cite{Anamiati:2016uxp} or the forward-backward asymmetry in their decays \cite{Hernandez:2018cgc}. In our model, the production through mixing with the SM neutrinos is in general suppressed by $yv/M \sim m_\nu/M$ and some extension like a left-right symmetric $SU(2)_L\times SU(2)_R$ model would required to have reasonable production cross section as discussed in \cite{Anamiati:2016uxp,Das:2017hmg,Dev:2019rxh}. 

While we discussed our model in the context of a mirror world, testable quasi-Dirac leptogenesis with the same features discussed in this work can also be realized in a minimal model where the mirror leptons and Higgs are gauge singlets. 
In this case, while one loses the attractive motivations of mirror world models, one can escape the associated cosmological problem of excessive dark radiation as the mirror particles could have low abundance if they are not excessively produced by the out-of equilibrium decays of heavy quasi-Dirac fermions which could also have small decay branching ratios to the mirror sector.\footnote{See for example Section 2 of \cite{Earl:2019wjw} and references therein for the discussions on the alleviating the dark radiation issue in the context of Twin Higgs models.} Therefore, a detection or limit on dark radiation will be a complementary test of mirror world model in general, quite independently of the quasi-Dirac leptogenesis we have proposed here.

\section{Conclusions}

We have proposed a simple model with small $B-L$ violation which is able to realize resonant leptogenesis at around weak scale as well as to produce quasi-Dirac mass spectrum for light neutrinos.
Remarkably, the parameter space for viable leptogenesis spans over the neutrino mass squared difference in the range $10^{-12}-10^{-6}\,{\rm eV}^2$ which can be probed in 
neutrino oscillation experiments, while for the case of even smaller mass splitting, be probed by flavor content of high energy neutrinos. 
Nature might have chosen quasi-Dirac spectrum for light neutrinos over purely Dirac or Majorana mass spectrum. This model, with its intimate connection to the cosmic baryon asymmetry, should serve as a prototype model to be fully explored in experimental searches.

\section*{Acknowledgments}
This work was supported in part by the Natural Sciences and Engineering Research Council of Canada (NSERC). AT would like to thanks T. Toma for providing the RG equations of the Majorana masses. C.S.F. acknowledges support by FAPESP, grant 2019/11197-6 and CNPq, grant 301271/2019-4.

\bibliographystyle{JHEP}
\bibliography{biblio}

\end{document}